\documentclass[conference]{IEEEtran}

\usepackage{cleveref}
\usepackage{listings}
\usepackage{courier}
\begin{document}
\title{Software Engineering For Automated Game Design}

\author{\IEEEauthorblockN{Michael Cook}
\IEEEauthorblockA{School of Electrical Engineering and Computer Science\\
Queen Mary University of London\\
mike@possibilityspace.org}
}

\maketitle

\begin{abstract}
As we develop more assistive and automated game design systems, the question of how these systems should be integrated into game development workflows, and how much adaptation may be required, becomes increasingly important. In this paper we explore the impact of software engineering decisions on the ability of an automated game design system to understand a game's codebase, generate new game code, and evaluate its work. We argue that a new approach to software engineering may be required in order for game developers to fully benefit from automated game designers.
\end{abstract}


\IEEEpeerreviewmaketitle

\section{Introduction}
Artificial intelligence has shaped the games industry for decades. In some cases this takes the form of player-facing, direct impact, such as providing intelligent controllers for in-game agents \cite{goap}. In other cases this impact occurs behind the scenes, by assisting in the development of games, improving the tools and processes that developers interact with every day \cite{speedtree}. The adoption of new AI techniques, however, takes time. For example, MCTS \cite{mcts} -- which rose to prominence in game AI research in the late 2000s and was instrumental in AlphaGo's defeat of Lee Sedol in 2015 \cite{alphago} -- has only been used in a handful of commercial AAA games.

There are many reasons for this. In \cite{tidd} Tidd and Trewhella identify two key factors affecting the adoption of new technology: `comfort' (how easy it is to integrate) and `credibility' (how likely it is to help the business). In terms of `comfort', games tend to be developed on extremely tight schedules -- 75\% of respondents to the 2019 IGDA Developer Satisfaction Survey report working overtime or `crunch', showing how game development is overstretched \cite{igdadss}. Thus, sparing workers to acquire this knowledge, or retraining and reshaping workflows to integrate it, is likely to cause delays and increase pressure on workers, at least in the short-term. In terms of `credibility', being the first to adopt a new technique is risky. If there are no or few examples of the technique working in a commercial context, the exact costs and benefits of using it in a commercial-scale project are unknown. This is one of the reasons academics face repeated calls to demonstrate their work in `real' games\footnote{Which is a longer discussion that this footnote is too small to contain.}.


Automated game design (AGD) research is the study and engineering of AI systems that actively participate in the design of games -- both through creating, critiquing and editing their core systems, and through creating content with an understanding and appreciation for how that content relates to the game design. AGD systems can act as autonomous or co-creative partners in the design process \cite{gom}, or take a more passive role by performing services on request \cite{sketchbook}. AGD has the potential to greatly impact areas of game development that have rarely been affected by AI up to this point. However, in terms of `comfort' and `credibility', AGD in its current form is unlikely to be adopted by the industry for some time.

To date, all AGD systems for digital games use \textit{game description languages} (GDLs), custom domain-specific languages for describing games. GDLs are usually interpreted by a custom engine, allowing the AGD system to write, edit and interpret simple high-level descriptions, while allowing them to be played and tested as a full game. This has many advantages: games are portable, because they are represented as text files rather than fully compiled binaries or folders of code; the design space is greatly reduced as the GDLs used are much less verbose than a modern programming language; and it makes collaboration between researchers easier by providing a way to describe games not tied to a language, engine or platform. VGDL is the best-known example of an academic GDL, which has been used by multiple researchers \cite{vgdl}.

Despite these advantages, GDLs may also be a barrier to AGD's long-term adoption by developers. Most modern game development practices, from professional to hobbyist, do not use the kind of high-level GDLs used by AGD systems. Tools such as Game Maker or Stencyl de-emphasise programming by using visual interfaces to help users describe games, while most other game developers (especially larger AAA studios) work directly in programming languages such as C\# and C++, possibly in conjunction with a development tool such as Unity or Unreal. The closest analogue to a GDL is perhaps Puzzlescript, a comparatively niche development tool which bears some similarities with VGDL, although its rules definitions are considerably more complex than VGDL's. Overcoming this difference is not simply a question of adapting research to industry -- building AGD systems that do not rely on GDLs is a different research problem, and one which raises new questions and opens new directions for AGD research. 

In this paper we investigate the challenges of building AGD systems that work directly with code, and envisage an AGD system designed to be integrated with a large, existing game codebase. Specifically, we explore how software engineering decisions made during development impact how an AGD system is able to analyse, generate and evaluate code. We do this by examining the problem at the low-level of individual lines of code and methods, and at the high-level of software design patterns and engine structures. We argue that certain patterns and approaches to writing software benefit code-based AGD systems, and that by identifying such patterns we might bring about a new paradigm for software engineering in which games are developed with a consideration for the needs of both human and AI game designers.


The remainder of the paper is organised as follows: in \ref{sec:background} we provide background on some aspects of software engineering relevant to our paper; in \cref{sec:accessibility,sec:typesigs,sec:postconditions,sec:patterns} we highlight several concepts in software engineering and explain how they impact the operation of automated game design systems; in \cref{sec:related} we discuss existing work which approaches code-based AGD, and how this may indicate future trends; finally, in \cref{sec:discussion} we explore issues emerging from this paper, and future research avenues this opens up.

\section{Background: Code Specification}\label{sec:background}
Code specification (whether for a full program or a single instruction) describes the expected behaviour of code, as well as any guarantees made about the output of the code, or constraints placed on inputs supplied to the code. Preconditions (logical statements which must hold before the execution) and postconditions (logical statements that are guaranteed to hold after execution) are examples of code specifications.

Code specification can be \textit{formal} or \textit{informal}. Formal specifications are usually mathematical or logical in nature, and sometimes can be implemented in code. For example, program logics such as Hoare Logic \cite{hoare} allow the precise specification of the behaviour of a command. Programming languages themselves are a kind of formal specification -- type systems, for example, are a way of formally describing constraints on data that is passed around and manipulated by code. While type systems have become a very commonplace example of formal specification, there are also examples of more unusual features which allow for even more formal specification, such as Haskell's Liquid Types which allow additional constraints to be attached to type declarations \cite{liquidhaskell}, or Rust's Ownership Types which constrain how threads can access a type, and whether they have read or write access \cite{rusttypes}.

\textit{Informal} specifications are usually expressed in everyday language, although this in itself varies in formality. Many programming languages have written documentation which describe instructions, keywords and library functions. Although such documentation is often written precisely, we consider it to be informal code specification because it is not mathematical or rigorous and can be ambiguous and open to interpretation, nor is it executable as part of a program. Certain areas of research are dedicated to studying and formalising the written specifications of programming languages and libraries \cite{raad}. Inline documentation (e.g. comments on code) are also informal specification, which may be more or less detailed depending on the programmer and any enforced code style. Less formal still are specifications which are not written at all -- the mental model constructed by a programmer as they write code, or interpret and use code written by someone else.

As code specifications decrease in formality, they are increasingly likely to be \textit{incomplete}, meaning they do not accurately describe every possible way the code can execute. Sometimes this can lead to errors, such as a mismatch between the written-language specification of code, and its actual function. Even when correct, less formal specifications can elide important details about the pre- or postconditions of a code block. As we demonstrate in the following sections, if an AGD system is to interact directly with a codebase, any element of the code specification that is not accessible through meta-programming (such as using reflection to examine the return type of a method) is inaccessible to the system. This is important because informal specifications are easier to read and write than formal specifications, which makes them more commonly used and as a result means that a lot of important specification details are not available to an AGD system. 

In many cases it is not possible to express code specification in a given programming language. In the popular Unity game engine, using C\# as a main project language, the most formal specification tool available to a programmer is its type system. Some additional specification tools exist, such as the ability to assert logical statements, but the specification expressed by these features cannot be accessed via metaprogramming, and often include informal elements (such as a written explanation providing context to the assertion).


\section{Software Engineering for Automated Game Design}
In the following sections we discuss several aspects of software engineering and design. In each case we introduce the concept, provide a concrete example in the context of a game codebase, and then illustrate how variations in implementation affect the way an AGD might interface with the code. The issues we discuss in this paper emerged from our work developing a new AGD system which integrates directly with existing codebases, and uses it to synthesise and evaluate new code to solve game design tasks \cite{cookcog20}. However, we present our discussion here without reference to a specific system, and instead discuss these emerging issues in the general case. 

Our general-case refers to what we call `code-based AGD systems'. Such systems are not defined by the AI techniques they use, or the area of game design they operate on. Instead, we define them as any AGD system whose main inputs and outputs are program code, rather than an abstract GDL representations. By \textit{input} we mean information that the system uses to generate and evaluate artefacts, and by \textit{output} we mean the artefacts produced by the design process. For example, in \cite{mechanicminer} Cook et al describe a system which uses metaprogramming to search a game's codebase in order to generate new game mechanics, which are output as Java code snippets that fit into the codebase. This would constitute a simple example of a code-based AGD system.

Common features of such systems would include reflection and dynamic code generation and execution. Reflection is the ability of a program to examine the active codebase at runtime, obtaining information such as lists of fields and methods in classes, and performing actions like invoking methods and changing variables. It is not necessary to understand these concepts for the following sections; simply assume that the AGD system in question can examine, extend and execute the game's codebase at runtime.

\section{Accessibility and Encapsulation}\label{sec:accessibility}
Accessibility modifiers affect the way in which classes and class members (e.g. fields and methods) can be accessed by different parts of a codebase. For example, in C\# a field with the \textit{private} modifier can only be accessed by code within the file it is defined in. This allows programmers to control how data is accessed across a project, or when a codebase is integrated with another. In C\# access modifiers are used at compile-time to verify that code throughout the project respects the stated access constraints.

From the perspective of an AGD system that works with code, reducing accessibility can be seen as an advantage in some cases as it reduces the size of the generative space that the system is working in. Most game codebases will include a lot of data in scope which is either not relevant to the task the AGD system is solving, or is only relevant in extreme cases. Such fields or methods complicate the generative task by increasing the options a code generator has to choose from with no equivalent gain in expressivity. Simply making the field inaccessible to the code generator by giving it a stricter accessibility requirement improves the chances that code generated by the AGD system will be useful.

A common programming pattern is to \textit{encapsulate} data such as class fields by giving them restrictive accessibility modifiers (such as \textit{private}) and to supply access methods instead, sometimes colloquially called \textit{getters} (for reading values) and \textit{setters} (for writing to them). This has many advantages, such as easily allowing refactoring of a variable without changing any of the code that references it. In general, reducing accessibility is seen as `safer', because it limits the extent to which data can be accessed or changed at runtime.

Though safe, encapsulation can complicate the generative space for an AGD system. Encapsulating a primitive type field such as an integer, for example, removes one field from scope and adds two methods (one getter, one setter) to it. Let us consider reading and writing with and without encapsulation, using the example of an integer field, \verb|x|, with getter \texttt{GetX()} and setter \texttt{SetX(int x)}.

In the case of reading, an AGD system might be interested in generate an integer expression (in order to pass to a method call, for example). In this case, there is little difference between the encapsulated and non-encapsulated case: both \texttt{GetX()} and the literal field access \texttt{x} are of integer type, and thus both are equally easy to discover unless the system distinguishes between method calls and variable reads.

In the case of writing, however, there is a distinct difference between the two cases. Writing to a variable (e.g. \texttt{x = 3}) is an \textit{assignment} operation which has a return type equal to the left-hand side (in this case, 3). Calling a setter method (e.g. \texttt{SetX(3)}) is a \textit{method invocation} and will usually have a return of type \texttt{void} (it will not return anything). This means that an AGD system will generate these expressions with different likelihoods under different conditions. For example, if the system is to generate an expression of type \texttt{int} it would be valid to generate the expression \texttt{x = 3}. However, it could not generate an invocation of \texttt{SetX(3)} though similar, because the expressions do not have the same type.

In addition to being generated under different circumstances, the two expressions provide different amounts of information to the AGD system, which might affect how well it can use the expressions when generated. The assignment expression is defined in the grammar, which means that when the system generates such an expression the functional specification of the code is known. By contrast, a method invocation provides some functional specifications (see \textit{Type Signatures}, below) but the postcondition of the method invocation, namely that a variable has had a new value written to it, is not stated. 

The reason for this is that encapsulation adds additional \textit{informal} specification at the expense of losing some \textit{formal} specification. The formal specification provided by the known properties of a field and its type are lost. On the other hand, the added informal specification cannot be accessed by the AGD system. For example, the naming convention of prefixing a setter method name with \texttt{Set} indicates to the programmer what the function of the method is, and may be accompanied by written documentation describing the purpose of the method. All of this constitutes informal specification, which is very useful for a programmer. However, this cannot be accessed or understood by metaprogramming, and thus this specification is not useful for an AGD system.

\section{Method Type Signatures}\label{sec:typesigs}
A method's \textit{type signature} is a partial specification that describes how a method can be invoked, what information can or must be passed to it, and what type its return value has. Consider the following type signature of a method for adding two numbers:

\begin{center} \texttt{public int Add(int x, int y)} \end{center}

The \texttt{public} keyword is an accessibility modifier, discussed above. The \texttt{int} type indicates that this method returns an integer. Finally, \texttt{Add} denotes the method name, followed by a sequence of arguments and their associated types. There are other more complex language features that can appear in C\# type signatures, including the \texttt{virtual} modifier, generic types, and modifiers such as \texttt{ref} and \texttt{out}. 

A method signature is a kind of formal specification. C\# verifies at compile time that invocations of this method subscribe to the rules laid out by its type signature -- that the arguments passed to it comply with the given types, for example, and that the result of invoking the method is treated as having the same type as the method's return type. Such information can be used by the AGD system to understand where it can appropriately invoke the method, and how to do so -- for example, knowing what expressions it must generate to pass as arguments.

There are many ways to write any given method, and different implementations will have certain advantages. Efficiency, generality, personal coding style, readability and ease of maintenance are just a few factors that affect how a particular piece of functionality is implemented in a codebase. An implementation may also be affected by working in a team -- some larger development teams may have an overall code style that they adhere to, for example. Figure \ref{fig:move_example} shows three different implementations of a \texttt{MoveObject} method, which updates the location of a game object.

In the first example the parameters passed are the new location. In this case the location passed is absolute; it does not depend on the object's current location. In the first second example, however, the parameters passed represent the \textit{change} in location; moving left subtracts from the current x co-ordinate, whereas moving right adds to it.  Finally, in the third example a special \textit{enumerated type} is used representing the cardinal compass directions, and the method uses these to move the object by one unit in the supplied direction.

Although these implementations are similar and the operation involved is quite simple, from the perspective of an AGD system they are quite different. For example, in a game where the game world is represented by a two-dimensional array (and therefore indexed by non-negative integers) the first example has an implicit, unstated precondition to avoid accessing the array with negative indices:

\begin{center} $x \geq 0 \land y \geq 0$ \end{center}

Similarly, in the case of a grid-based puzzle game in which an object can only move one grid space at a time, the second example has an different unstated precondition:

\begin{center} $-1 \leq \verb|dx| \leq 1 \land -1 \leq \verb|dy| \leq 1$ \end{center}

In both cases there are additional preconditions related to not moving out of grid bounds with too high a location value, which we omit here for space. Note that although the acceptable range of values for the first example's arguments are much smaller than the second example, the method signature for both examples is identical -- two integers, with a void return type. If an AGD system were to invoke either of these methods, it has no information suggesting how it should narrow its input range. This means that it is quite likely to generate code that causes runtime errors, especially if it is allowed to generate literal expressions (i.e. if it is allowed to pass arbitrary numbers as arguments to a method).

Let us now consider the third example. Although this method has similar implied conditions on movement off the edge of the grid (for example, moving west is not allowed if the x co-ordinate is 0), its range of legal inputs is much smaller than the other two examples, because its arguments are expressed in terms of a custom type with just eight values. The custom type adds additional formal specification to the definition of the method. When an AGD system generates an invocation of this method, it is far less likely to generate arguments which are not legal. 

However, this comes at a tradeoff. For example, in the first two cases it is possible to move an object by more than one grid space with a single invocation of \texttt{MoveObject}. Even if it is not intended to be in the game, such a mechanic might be something that would be desirable for an AGD system to invent or discover. In the third case, this would only be possible by invoking the method multiple times with the same arguments -- not only is this less efficient, it is less likely to be generated by a system which synthesises code line-by-line.

Additionally, such custom types reduce the chance for experimental combinations. For example, suppose an AGD system generates a new player ability which invokes \texttt{MoveObject}. In the first two examples where the arguments are of type integer, there are likely many fields, methods and literal values the system could try passing as an argument. For instance, it might pass the player's current health as an argument, creating an ability that lets the player move further the less damage they have taken. This is an unusual ability that the system could discover by combining different kinds of data of the same type. By contrast, in the third example the system is less likely to find existing fields or methods of type \texttt{DIR}, because the purpose of this custom type was to formally describe a particular kind of specialised data relating to direction. Thus, invocations of \texttt{MoveObject} are safer, but less likely to lead to surprising results.


\begin{figure}
\begin{lstlisting}
//Implementation 1
public void MoveObject(int x, int y){
	this.x = x;
	this.y = y;
}
//Implementation 2
public void MoveObject(int dx, int dy){
	this.x += dx;
	this.y += dy;
}
//Implementation 3
public enum DIR {N, NE, E, SE, S, SW, W, NW};
public void MoveObject(DIR direction){
	switch(direction){
		case N: this.y += 1; return;
		// cut for brevity
	}
}
\end{lstlisting}
\caption{Three implementations of a method which moves a game object.}
\label{fig:move_example}
\end{figure}


\section{Postconditions and Side Effects}\label{sec:postconditions}
A postcondition is a partial code specification for a method or block of code, and describes a set of properties which hold after the code has finished executing. For example, the return type of a method constitutes a postcondition describing the type of the returned value. Postconditions can also be more elaborate, describing the functionality of the code. For example, a list-sorting algorithm has a postcondition expressing how the returned list will be ordered. As with other kinds of code specification, postconditions may be expressed formally; written in comments or reports; understood informally by the programmer; or not expressed at all. 

A \textit{side effect} of a piece of code refers to any effect caused by executing it that is not part of its return value, and that persists after the code has finished executing. This can be thought of as any modification of the heap that outlasts the code's execution. For example, all of the examples in \cref{fig:move_example} modify the fields \texttt{this.x} and \texttt{this.y}, and are therefore side effects of the method's execution. 

Both postconditions and side effects pose a problem for AGD systems that work with code directly. Postconditions are rarely formally specified beyond stating the return type of a method. Based on our interviews with game developers, and the authors' own experience making games, elaborate postconditions are at most expressed through informal written documentation. This information cannot be accessed by an AGD system. These more elaborate postconditions are critical to reasoning about control flow and the wider structure of a program, and without this information AGD systems can only invoke methods based on their return type.

Similarly, side effects are almost exclusively expressed through informal written specification. Side effects are common in imperative programming languages such as C\#, because methods are often written specifically for this purpose rather than transforming inputs. This causes two problems for an AGD system. The first problem is that \textit{intentional} method invocation becomes difficult: since the system cannot know the purpose of, or existence of, method side effects it must invoke them blindly, without knowing what they are intended to be used for. Consequently, the second problem is that \textit{unintentional} side effects as the result of executing code are more common. An AGD system might invoke a method in order to generate an expression of a particular type (based on its return value) without realising that the method invocation has additional side effects. 




\section{Software Design Patterns}\label{sec:patterns}
Design patterns are high-level approaches to structuring entire codebases in order to encourage certain ways of writing code, controlling data, or connecting information. Design patterns are often developed to solve commonly recurring problems in development. For example, \textit{object pooling} is a pattern which allows for a small collection of objects to be continuously reused within an application, to avoid expensive repeated instantiation. Game developers often use a wide variety of software design patterns, whether working alone or in large teams \cite{gamepatterns}. In this section we describe two examples of a software design pattern, and show how they can have a large impact on code-based AGD systems.


\subsection{Model View Controller}
The Model View Controller (MVC) pattern is a software design pattern that separates an application into three distinct areas of responsibility: the \textit{model}, which defines the logic and functionality of the application; the \textit{view}, which defines how the application's current state is rendered; and the \textit{controller}, which defines how a user can interact with and change that state \cite{mvc}. This separation of concerns decouples two important and highly variable parts of an applications (rendering and input) from the less-variable core, which is very useful for game development when an application is built for multiple technology platforms, input devices, resolutions and users.

While the MVC pattern is useful for software developers in terms of clear structure, the pattern also necessarily distributes the code for important functionality to separate parts of the codebase. The consequence of this is that the creation of a new game object may require intervening in multiple areas of the code. For example, adding a new item to an action-adventure game will involve describing how to render the item and any associated effects (in the \textit{view}), describing what part of the control scheme accesses and uses the items (in the \textit{controller}), and adding code describing the effects the item has on the game's systems (in the \textit{model}). 

The decentralisation of code caused by patterns like MVC are natural for people to think in terms of, but complicate AGD work by requiring multiple changes to the codebase in order to add certain kinds of content (such as the new item example above). It is possible to circumvent these problems with additional bootstrapping from a developer, for example by separating a design problem into subproblems which relate to the model, view and controller components. This increases the burden on the developer slightly, but perhaps more importantly also increases the knowledge required by the developer to work with the AGD system. This might limit who is able to work with such a system, if for example they are not familiar with the structure of the codebase.



\subsection{Entity-Component Systems}
The Entity-Component System (ECS) is a software design pattern that constructs objects (or \textit{entities}) out of one or more \textit{components}, smaller classes that represent a particular capability or property \cite{ecs}. A component is isolated from the larger entity it is a part of, and communicates instead by broadcasting messages which other components may receive, respond to, modify or ignore. For example, a monster might send a \texttt{ReceiveAttack} message to the player in an RPG, with a damage value. The message is received by the \texttt{Armour} component in the player entity, which modifies the damage value by half and passes the message on. The message is then received by the \texttt{Health} component which subtracts the damage from the player's health and prints a message.

ECS emphasises \textit{composition} over \textit{inheritance} -- it encourages the reuse of code, but not through direct subclassing. This allows for ontologies with complex behaviour reuse, and as a result ECS has become particularly popular with developers of highly systemic or simulation-driven games, with notable examples including \textit{Dwarf Fortress} and \textit{Caves of Qud}. ECS is also popular in AAA game development, and many tools like Unity use ECS as the basis for their engine. 

ECS is an example of a design pattern that has benefits for AGD systems. In fact, we believe ECS-driven games such as roguelikes may be the best format with which to explore code-based AGD research initially. There are two main advantages ECS provides. First, the structure of game objects is simple, robust and clear. In many ECS implementations an object has no code of its own, and is simply defined as a list of components. Since components are self-contained, an object does not need to decide about control flow or method invocation -- at most, it may be required to set component fields to certain starting values. This makes generating, comparing and recombining objects very straightforward.

Second, the invention of new components is a more tractable problem than the open-ended task of general code block design, because components can only communicate with the codebase by sending and receiving messages. The available message types are often formally specified within the codebase through enumerated types, and usually only carry small amounts of data. In addition to being simple and lightweight, messages are also less likely to violate code specification compared to method invocation, because the default response to a message is to ignore it. Thus, if a message is inappropriate to send to an object or component, the most likely outcome is that nothing happens.

ECS implementations vary, and not all uses of the pattern will have the properties we describe above. However, by enforcing a very simple code style (for example, requiring that all messages are described in an enumerated type) we believe an ECS driven-game is a perfect foundation on which to build code-based AGD systems. This is especially encouraging given that many common uses of ECS are for games with high degrees of emergence and systemic complexity, which is an ideal environment for an AGD system to work in and potentially contribute novel and interesting design ideas.

\section{Related Work}\label{sec:related}

In \cite{variationsforever} the authors present \textit{Variations Forever}, a system which procedurally generates micro-games. Their overall aim was to allow the player control over the rulesets generated for these micro-games, in the style of \textit{Endless Web} \cite{endlessweb}. Variations Forever is notable here for its use of answer set programming (ASP) to define a rules space through constraints. While ASP is a fairly complex programming paradigm that is not widely adopted, the concept of interacting with an intelligent designer by expressing limits and constraints is very appealing, as it allows designers to express requirements without restricting a system's exploration of the rest of the design space. We anticipate that code-based AGD systems will allow for easy definition of which parts of the codebase the system can access and use, probably through language features like annotations.

In \cite{mechanicminer} the authors describe a system which can generate new single-button game mechanics that help the player solve a problem in a puzzle platformer. The game mechanics are generated by searching the game's codebase using metaprogramming to find variables that can be changed or toggled. It then evaluates these mechanics by attempting to solve a small challenge level which is considered unsolvable using the game's basic mechanics. If the challenge becomes solvable, the system infers that the generated mechanic must allow the player to change the game's state to their advantage. This technique of testing generated code against small test cases with guaranteed binary properties (i.e. solvable/unsolvable) is, we believe, a clear and effective way of evaluating code for certain types of game content. We expect to see these kinds of `gameplay unit tests' commonly used to make it easier for designers to express evaluation conditions for AGD systems.

In \cite{gemini} the authors describe Gemini, an AGD system which uses a specialised DSL called Cygnus. A notable feature of Gemini is its ability to work bidirectionally -- it is able to generate games from descriptions, and also able to offer interpretations of games. A major problem for AGD systems is contextualising their work \cite{cookaiga}, and this is amplified when working as a contributor to a much larger game project. Gemini shows how a system might be able to create explanations of created artefacts, that potentially understand the broader goals and creative direction of the game it is attached to. This might involve building an interpretation of the existing codebase before creating new content, and then finding a way to reconcile the generated design content with that interpretation after the fact.

\section{Discussion}\label{sec:discussion}
\subsection{A Need For Formal Specification}
A common theme throughout many of our examples is a tendency for people to rely on informal approaches, such as written documentation or naming conventions, to convey code specification. In many cases this results in code with less formal specification, but an overall stronger structure for a programmer. For example, in the case of encapsulation and accessibility, getters and setters are considered safer than public variable access. With an understanding of this convention, a programmer benefits from a safer codebase with little impact to their workflow. However, an AGD system cannot access informal specifications, which means that although these approaches are better for people, they are worse for automated analysis.

Programming language research is constantly developing new tools and techniques to make it easier for programmers to write correct, well-specified and efficient code. The history of research in this area has given rise to type systems, language annotations, and assertion statements. However, the adoption of these ideas are slow, and the net benefits of formal specification may not be clear. Modern programming practices have ways to ensure code correctness, and these rely highly on informal specification and human interaction -- code review, for example. More formal approaches are less common. While we have showed that code specification can be conveyed to an AGD system through changes in the structure of code, we should also consider simple ways for game developers to provide small amounts of formal specification for an AGD system to use. In \cite{cookcog20} we propose annotations as a good starting point for this, as it is already a standard language feature in many common programming languages.


\subsection{Tractability Versus Surprise}
A common tradeoff in generative and creative systems is between control over the system's generative space, and the scope for surprise and novelty from the system. By constraining the generative space we may be able to efficiently remove large amounts of poor-quality outputs, but in doing so we often remove areas of the space that contain interesting, unusual or creative artefacts as well. The use of game description languages (GDLs), which we discussed in the introduction to this paper, is another example of such a tradeoff. GDLs provide control over what exactly an AGD system can change within a game, but as a result they greatly restrict the kind of games it can produce (often within well-explored design spaces like arcade games). By contrast, being able to read, edit and generate code expands the generative space to contain a vast quantity of new and innovative games, but the size of the generative space increases by such a huge factor that the ratio of good games to bad becomes vanishingly small.

Some of the examples discussed in the previous section contain suggested solutions which help provide more information to an AGD system by limiting the space it works in. For instance, in section \ref{sec:typesigs} we gave an example of using a custom enumerated type, rather than an integer, to describe the direction an object moves in. In doing so, we make the space the AGD system works in more tractable, reducing the space of possible arguments from any valid integer value -- potentially $2^64$ values -- to just eight. However, we also limit how \texttt{MoveObject} can be called, as it can no longer be passed arbitrary integers. This removes the possibility of moving more than one space at a time, teleporting, having movement be affected by variables such as speed, or any number of other innovative uses of the method.

One of the advantages of an AGD system working directly with code is that it can connect unusual parts of a game's codebase together. In our prior work in \cite{mechanicminer}, our AGD system invented game mechanics that let the player control the game's physics system, allowing them to change the bounciness and gravity of the level at will. This novel connection of two disparate parts of the game's codebase results in something interesting and novel, but is only possible because of a lack of constraints. This tradeoff is a key tension in the design of code-based AGD systems, and will likely be a large area of future study and work. We believe that it is not necessarily up to us to decide on this tradeoff in advance; rather, it is important to make it easy for developers to control this tradeoff themselves.

%
%
%
%
%

\subsection{Automated Discovery of Code Specifications}
Inferring the specification of a piece of code automatically is an open research question. Many tools have been developed to tackle the problem, such as Facebook's Infer \cite{infer}, which focuses on memory safety issues. Inferring the \textit{functional} specification of a piece of code -- i.e. what it \textit{does} -- is a less explored problem in this field, as it is a much harder task. Some approaches use abduction in order to logically conclude properties about the code, while other approaches use sampling and testing to enable probabilistic inference.

The automatic discovery of functional specifications is a problem that is unlikely to be solved in the near future, and to our knowledge has not been attempted for code that works in creative domains such as videogame design. Nevertheless, we may be able to take inspiration from existing research, as well as building small tools and extensions to focus on extracting certain kinds of information. We have done some early experimentation with a similar problem in a tool for automatically analysing procedural generators, in which we attempt to automatically estimate a safe upper and lower bound for parameters by repeatedly testing different values and catching errors \cite{cook19cog}. 

Even simple features such as bounds estimations for parameters could greatly improve the ability of an automated game design system to use methods and fields in a more effective way -- understanding, for instance, that the position of an object can never exceed the dimensions of the array representing the world. As we have shown in this paper, even small amounts of information about the useful ranges for input arguments or other simple restrictions on method calls can drastically reduce the chance of generating non-compiling or exception-throwing code, and focus the search on more useful subsets of the generative space.

\subsection{Experimental Programming Language Features}
Our motivation for this work was to explore code-based AGD systems that could be useful to game developers without requiring large changes in their everyday game development practice. However there is extensive research into new programming paradigms, features and tools that could help us achieve richer in-code interactions between an automated game designer and a user. One example of this is \textit{refinement types}, introduced in \cite{refinementtypes} by Freeman and Pfenning. A refinement type is a type with an attached predicate that must always hold for any element of that type. Thus a refinement type for an integer might restrict the integer to hold a non-negative value. The intention behind refinement types is to enhance the code specification already provided by type systems, as a way to detect more errors at compile-time by formally expressing tighter constraints on types in different parts of a program. 

Throughout this paper we discuss how an increasing reliance on informal specification poses a problem for AGD systems. On possible solution to this, which we also explore in \cite{cookcog20}, is the use of annotations to provide additional information that contextualises methods and fields. A full refinement types implementation would allow for a much wider variety of predicates to be attached to typed expressions. However, refinement types have seen little use in programming, with Haskell's Liquid Types being the only major implementation in a popular language \cite{liquidhaskell}. It might be possible to implement a version of refinement types specifically for a particular game engine, such as Unity, to provide more control over an automated game designer.

\subsection{ECS Roguelikes -- A Next Step}
As we discussed in \cref{sec:patterns}, although code-based AGD is an intimidating research problem, some game structures and formats may be more tractable than others. In particular, we believe that entity-component systems offer a good balance between code-based AGD and GDL-based AGD. They allow us to write code-based AGD systems that work directly with game code, while also benefitting from a tight code structure, well-defined objects and components, and a simple and robust message-passing system. We intend to explore this as a next step in the area of code-based AGD, and we hope to see other AGD researchers do the same.

\section{Conclusion}
In this paper we have presented an analysis of the impact of software engineering decisions on code-based automated game designers. We have shown that decisions about individual lines of code, as well as high-level decisions about the structure of entire projects, both have significant impacts on how well an AGD system is able to understand, generate and evaluate code. Although this paper only has space to contain a few such examples, we have already identified several more, and we anticipate that in the future it will be possible to compile a set of best practices for developing games with AGD collaboration in mind. These techniques would vary in effort required, and in the benefits received, and would likely develop over time as code-based AGD systems grow in complexity and we better understand how designers use wish to use them.

The arguments we have laid out here point to new research questions for AGD researchers, particularly regarding how users will interact with AGD systems that are acting as partners in the development process, and whether it is possible to automatically analyse code and identify areas that could be changed to improve readability for an AGD system. This area of study is important, in our opinion, not just because it furthers our understanding of AGD systems but also because it explicitly frames AGD research as something complementary to, rather than a replacement for, people. In order to ensure that AGD research is used to augment and support the skills of existing game developers, we must perform our research with this in mind. The work in this paper, and the systems we are building alongside this, assume that AGD systems work in concert with people to achieve their goals. We believe that making this the default assumption in AGD research is an important part of setting expectations in the games industry and beyond, regarding the impact of automation.





%

\bibliographystyle{IEEEtran}
\bibliographystyle{IEEETran}
\bibliography{biblio}
%
%

\end{document}